\documentclass[12pt,preprint]{article}
\usepackage{epsf}
\newcommand{\eqb}{\begin{equation}}
\newcommand{\eqe}{\end{equation}}
\newcommand{\dmb}{\begin{displaymath}}
\newcommand{\dme}{\end{displaymath}}
\newcommand{\pd}{\partial}

\newcommand{\eab}{\begin{eqnarray}}
\newcommand{\eae}{\end{eqnarray}}

\newcommand{\e}{\mbox{e}}
\newcommand{\be}{\begin{equation}}
\newcommand{\ee}{\end{equation}}

\newcommand{\La}{\Lambda}

\begin{document}
\begin{titlepage}
\begin{flushright}
MPI-PhT 2002-01 \\
\end{flushright}
\vspace{0.6cm}

\begin{center}
\Large{{\bf  A Model for the Effective Potential 
of Thermalized Pure SU($N$) Gauge Theory}}

\vspace{1cm}

Ralf Hofmann

\end{center}
\vspace{0.3cm}

\begin{center}
{\em Max-Planck-Institut f\"ur Physik\\ 
Werner-Heisenberg-Institut\\ 
F\"ohringer Ring 6, 80805 M\"unchen\\ 
Germany}
\end{center}
\vspace{0.5cm}

\begin{abstract}

Based on the (in part verified) ideas of a dynamical ``abelian-ization'' and subsequent 
``center-ization'' of pure SU($N$) gauge theory 
an effective potential for relevant field variables is constructed in the limit of large N. 
To do this the theory is assumed to be thermalized and to be 
gravitationally deformed. BPS saturation in the 
dynamics of the monopole field is shown to lead to a suppression 
of the back reaction due to classical gravity. 
The classical, effective description of the 
gauge theory can be justified for both 
the regime of maximal abelian gauge symmetry 
and the center symmetric phase.

\end{abstract} 

\end{titlepage}

\section{Introduction}

Employing loop-expansion within (resummed) perturbation theory, the calculation of effective potentials due 
to fundamental, renormalizable interactions 
has a long standing history \cite{ColemanWeinberg}. 
Applied to the framework of (imaginary time) 
finite temperature theory this 
approach has had its successes investigating the strength of phase 
transitions in the Abelian Higgs model and the electroweak symmetry breaking 
in the standard model \cite{SMT}. A common feature of 
the perturbative results is the fact that the temperature dependence of 
the ground state of the theory is always contained in the {\sl parameters} 
of the effective potential 
and {\sl not} in the {\sl solution} of the corresponding 
equation of motion. This is due to the fact that the ground state is, 
for reasons of calculational feasiblity, 
assumed to be a configuration of spatially {\sl constant} and 
temperature independent, gauge invariant, and scalar field strength. In the linear sigma model, 
which is a way to picture chiral symmetry breaking, finite temperature effective potentials 
were obtained by clearly abusing perturbation theory since there the expansion parameter is 
much larger than unity. Even the results of the afore mentioned gauge theory calculations 
cast doubt on the convergence of the perturbative expansion. A property of 
the perturbative treatment is the gauge invariance to each and every 
order in the coupling constant. This invariance 
is an important guiding principle for 
the organization of diagrams. Therefore it must not be violated - after all it is 
crucial to prove renormalizability of the theory. 
However, there are fairly strong indications that fundamental 
gauge symmetries masquerade as smaller gauge or even 
discrete symmetries at low energy \cite{KraussWilczek,center}. These subgroups 
are thought to be 
represented by composite degrees of freedom whose 
very occurence as vacuum dominating fields 
breaks the respective original symmetry spontaneously. Typically, one would have 
an adjoint, composite scalar field in pure SU(2) Yang-Mills theory whose condensation 
would generate magnetic monopoles and spontaneously would break the fundamental gauge symmetry down to the  
abelian subgroup U(1). The condensation of monopoles, in turn, would spontaneously break 
the U(1) gauge symmetry which in a final transition masquerades as the center symmetry $Z_2$ 
to render the theory in its confining phase. 
Viewed as a composite, the effects of the Higgs field in 
the Standard Model may have originated from a spontaneous break-down of 
a higher than SU(2)$\times$U(1)$\times$SU(3) gauge theory with no fundamental, 
scalar matter \cite{Susskind}.   

These cascading spontaneous symmetry breakings can not be captured in a perturbative approach. 
In spiritual analogy to Chiral Perturbation Theory ($\chi PT$) 
\cite{LeutwylerGasser}, which is non-renormalizable in the sense that at a fixed order in the momentum expansion 
there are divergences that can not be 
swallowed by a counterterm contained in the 
structure of the classical Lagrangian, we therefore propose an alternative 
to perturbative calculations of effective potentials at finite temperature.
We specialize to the case of SU($N$) gauge theories. 
The idea is that once the relevant 
degrees of freedom of a low-energy description are identified their 
interaction can be constrained by a small set of principles. 
In $\chi PT$ the only experimentally justified and very constraining 
assumption about the nature of pionic degrees of freedom 
is that they are the Goldstone bosons of a 
spontaneously broken, global symmetry. 

\section{General ideas}

Here we appeal to the intuition \cite{dualsc} that fundamental and 
pure SU($N$) gauge theories 
essentially masquerade as (dual) U(1)$^{N-1}$ Higgs models within some intermediate range of 
resolution. Thereby, a non-zero Higgs field expectation signals the condensation of 
correspondingly charged magnetic monopoles. These theories, in turn, may be invariant only 
under the discrete subgroups $Z_N$ as one tunes down the resolution even 
further \cite{KraussWilczek,center}. The confinement picture relying on the condensation of center vortices 
has been impressibly verified on the lattice for $N=2$ \cite{centerlat}. 
Note that there monopole and $Z_2$ vortex condensation 
were found {\sl not} to be mutually exclusive. On the contrary, 
the appearance of $Z_2$ vortices in 
the confining phase seems to imply a strong form of monopole condensation. Consequently, although an effective description 
may be in terms of scalar monopole fields, a transition to the phase, 
where only the center symmetry survives, is captured.  
A question worth asking then is whether there are, in analogy to $\chi PT$, 
physically motivated constraints so as to 
render the interaction of these Higgs fields unique. In this paper we argue that 
at least for large $N$ the answer seems to be yes.  

For the case $N=3$ it was argued 
in \cite{Suzuki} that the maximal abelian subgroup U(1)$^{3-1}$ can be 
promoted to U(1)$^{3}$ and afterwards diminished back to U(1)$^{3-1}$ by imposing one constraint 
on the sum of phases of the 3 Higgs fields and by (trivially) integrating out the 
additional gauge field. This is suggested by the observation that the 
magnetic charges of the dual U(1)$^3$ theory are integer multiples of the positive
(artificially extended) root vectors of SU(3). 
If this procedure would hold in general, one would have to impose \# (positive) 
roots=$\frac{1}{2}(N^2$--$N$) {\sl minus} \# physical magnetic charges=\# non-trivial 
center elements=$N$--1 {\sl equals} $\frac{1}{2}N^2$--$\frac{3}{2}N$+1 \# constraints on the \# in part artificially 
introduced Higgs fields=$\frac{1}{2}(N^2$--$N$). This, however, would make 
general considerations impractical. Therefore, one is forced to consider small $N$ 
for an exact treatment. Thus we choose to {\sl simulate} the case $N>3$ by considering a U(1) symmetry 
reducing to the center symmetry $Z_N$ at low temperature.

\section{BPS saturation and gravitational interaction}

It has proven useful in the past 
to consider interaction with gravity as a guide to derive justified 
expressions for, say, the energy-momentum-tensor in the limit of a flat but bounded 
spacetime \cite{CCJ}. We therefore start with the following action of an  
abelian Higgs model which is {\sl minimally} coupled to gravity
\eqb
\label{act}
S=\int d^4x \sqrt{-g}\left[
- \frac{1}{16\pi G}R - \frac{1}{4}F_{\mu\nu}F^{\mu\nu} + \overline{{\cal D}_\mu \phi}{\cal D}^\mu \phi - 
V(\bar\phi \phi)\right]\, .
\eqe
Thereby, ${\cal D}_\mu\equiv\pd_\mu+ie A_\mu$ denotes the gauge 
covariant derivative, and $V$ is the to-be-constructed 
effective potential for the Higgs field. 

In view of the thermal treatment, which we will apply below to the effective SU($N$) dynamics, 
we consider the theory with a {\sl Euclidean} signature of the metric with spacetime having the topology of 
a 4-torus so that in appropriately chosen coordinates the fields $\phi$ and $\bar\phi$ are 
periodic along each coordinate. Let us for the moment ignore the gauge 
sector of the theory. Our goal is to construct the potential $V(\bar\phi \phi)$ such that 
there exist solutions to 
\eqb
\label{eom2}
\nabla_\mu\nabla^\mu\phi=\frac{\pd V}{\pd \bar{\phi}}\ ,\ \ \ \ \ \ \ \ 
\nabla_\mu\nabla^\mu\bar{\phi}=\frac{\pd V}{\pd \phi}\ ,
\eqe
which can be obtained by ignoring an appropriately constrained 
gravity. Here $\nabla_\mu$ denotes the covariant derivative with respect to the 
gravitational background. The key is to look for 
BPS saturated \cite{BPS} solutions to the field equations (\ref{eom2}). 

Let us give an example. For BPS saturation along a compact dimension it 
is essential that the ``square root'' ${V}^{1/2}(\phi)$ of the potential 
possesses a single pole \cite{LDS}. So let us consider
\eqb
V(\bar{\phi},\phi)=\frac{\Lambda^6}{\bar\phi\phi}\ ,
\eqe
where $\Lambda$ is a mass scale. It is easily checked (see also \cite{HofKeil}) in general 
that fields $\phi,\bar{\phi}$ satisfying 
\eqb 
\label{BPSex}
\pd^\mu\phi=\frac{1}{2}\,\bar{V}^{1/2}(\bar{\phi})\ ,\ \ 
\ \pd^\mu\bar{\phi}=\frac{1}{2}\,{V}^{1/2}(\phi)\ ,\ \ \ (\mu=1,\cdots,4)\ ,
\eqe
do satisfy 
\eqb
\label{incompl}
\nabla_\mu\nabla^\mu\phi=\frac{\pd V}{\pd \bar{\phi}}+
\frac{\bar{V}^{1/2}(\bar{\phi})}{4}\sum_\kappa g^{\mu\nu}\pd_\kappa g_{\mu\nu}
,\ \ \nabla_\mu\nabla^\mu\bar{\phi}=\frac{\pd V}{\pd \phi}+
\frac{V^{1/2}(\phi)}{4}\sum_\kappa g^{\mu\nu}\pd_\kappa g_{\mu\nu}\ .
\eqe
Here $g_{\mu\nu}$ denotes the metric and $V\equiv\bar{V}^{1/2}(\bar{\phi}){V}^{1/2}(\phi)$, where ${V}^{1/2}$ is 
only determined up to a constant phase 
${V}^{1/2}_{\tiny{\mbox{real parameter}}}\rightarrow\e^{i\delta}{V}^{1/2}_{\tiny{\mbox{real parameter}}}$. 
For (\ref{eom2}) and (\ref{incompl}) to be the same 
we need $\sum_\kappa g^{\mu\nu}\pd_\kappa g_{\mu\nu}=0$. This would be the case if 
$g_{\mu\nu}=g_{\mu\nu}(s,t,u)$, where $s\equiv x^1-x^2+x^3-x^4,\,t\equiv x^1-x^2-x^3+x^4$, and  
$u\equiv x^1+x^2-x^3-x^4$. So the demand that the solutions $\phi,\bar{\phi}$ to the 
gravitation-free BPS equations (\ref{BPSex}) be also solutions to the 
second-order equations (\ref{eom2}), which involve a gravitational coupling term, 
implies that $g_{\mu\nu}$ depends at most on 3 rather than 4 coordinates. 
We hope to be able to give an example for a {\sl dynamical} 
$g_{\mu\nu}(s,t,u)$, which is periodic in $x^1,x^2,$ and $x^3$, in a future publication. 
For now let us only find periodic solutions to (\ref{BPSex}). 
Choosing $\delta=\pm\frac{\pi}{2}$ and factorizing the coordinate 
dependences,
\eqb
\phi(x)=\Pi_{\mu=1}^4 \phi_\mu(x_\mu)\ ,\ \ \ 
\ \bar{\phi}(x)=\Pi_{\mu=1}^4 \bar{\phi}_\mu(x_\mu)\ ,
\eqe
one obtains the following solutions 
\eqb
\label{solex}
\phi_\mu^{(n(\mu))}(x_\mu)=\sqrt{\frac{\La^3\beta(\mu)}{4\left[\Pi_{\mu\not=\lambda}\alpha(\lambda)\right]|n(\mu)|\pi}}\,
\e^{2n(\mu)\pi i\frac{x_\mu}{\beta(\mu)}}\ ,\ \ \ \ \ (n(\mu)\in{\bf Z})\ ,
\eqe
where $\alpha(\mu)$ is a constant of mass dimension 1/2 satisfying
\eqb
\label{alphas}
\alpha(\mu)=\frac{\Lambda^3\beta(\mu)}{4\left[\Pi_{\mu\not=\lambda}\alpha(\lambda)\right]|n(\mu)|\pi}\ ,
\eqe
and $\beta(\mu)$ is the $\mu$th cycle of the torus obeying
\eqb 
\label{betas}
\frac{\beta(\mu)}{n(\mu)}=\frac{\beta(\lambda)}{n(\lambda)}\ ,\ \ \ \ (\mu,\lambda=1,\cdots,4)\ .
\eqe
Note that the dependences of $\phi,\bar{\phi}$ on $\alpha(\mu)$ drop out thanks to (\ref{alphas}). 
A generalization of this 4-dimensional example to a spacetime of any even dimension and 
positive-definite metric is straightforward.

\section{The case of finite temperature}

Let us now return to the effective description of thermalized SU($N$) gauge theories. 
In general we have only one 
compact dimension here, namely $0\le\tau\equiv x_4\le\beta\equiv\frac{1}{T})$ where $T$ denotes the temperature. 
As for gravity we assume the Euclidean version of the Robertson-Walker metric although this is not essential. 
We construct $V$ by satisfying the following three constraints:\vspace{0.2cm}\\ 
1) The gravitational deformation of the ground state 
be encoded in the shape of the potential to a good approximation.\vspace{0.2cm}\\ 
2) According to what was said in section 2 
we the potential $V$ shall show a smooth transition to a $Z_{N}$ symmetric phase for field modulus $|\phi|$ 
close to a single mass scale $\Lambda$.\vspace{0.2cm}\\ 
3) The $T$ dependence of the ground state be entirely absorbed into the corresponding 
{\sl solutions} to the field equations and {\sl not} in the usual, perturbatively obtained 
$T$ dependence of {\sl parameters} of the effective potential. \vspace{0.2cm}\\ 
We will justify a posteriori that, except for the critical region where they drive the phase transition, 
quantum fluctuations are integrated out for large $N$ and hence are contained 
in the shape of the potential. Given this fact, the theory 
can be viewed as a classical one. 

An explanation of points 1)--3) is in order now. Point 1) is a 
call for simplicity. Rather than solving a fully coupled system of equations 1) 
ensures a partial decoupling. Applied to inflationary cosmology 
the implementation of this point causes a Hubble parameter which is 
much smaller than the mass of inflaton fluctuations 
during the de Sitter regime \cite{HofKeil}. This is in contrast to the usual slow-roll 
paradigm which implies fluctuations of small mass (compared to the Hubble parameter) 
during inflation. Point 2) incorporates the simplest possible dimensional transmutation: 
A dimensionless coupling constant $g$ plus $N^2$--1 fields in the fundamental theory 
are mapped onto a dimensionless coupling constant $e$, 2($N$--1) fields, 
plus a mass scale $\Lambda$ in the effective theory. Since we are only interested in 
global properties of the ground state a single mass scale is reasonable. 
Note, however, that disturbing the vacuum of SU($N$) gauge theory locally, 
a wealth of mass scales is to be considered \cite{Hof4}.

Point 3) is based on the idea 
that a $T$ {\sl independent} potential, which 
describes the thermal properties of the ground state in terms of the corresponding 
solution, may be of qualitative relevance in nonequilibrium situations where the mean 
resolution is considered an adjustable external quantity.  
  
In an application to very early cosmology it was shown in ref.\,\cite{HofKeil} 
that $\phi$ and $\bar{\phi}$ being 
solutions of the BPS equations, 
\eqb 
\label{BPS}
\pd_\tau\phi=\bar{V}^{1/2}\,,\ \ \ \ \ \ \ \ \pd_\tau\bar{\phi}={V}^{1/2}\ ,
\eqe
implies that away from Planckian initial conditions they are approximate solutions to the corresponding 
Euclidean second-order equations involving a small gravitational coupling term. 

Again, the existence of BPS saturated 
solutions along a compact dimension ($0\le\tau\le\beta$) necessitates 
$\bar{V}^{1/2}$ and ${V}^{1/2}$ to possess single poles \cite{LDS}. Together with 2) this fixes 
the potential uniquely as
\eqb
\label{pot}
V(\bar{\phi}\phi)=\frac{\La^6}
{\bar{\phi}\phi}+\lambda^2 \La^{-2(N-3)}
(\bar{\phi}\phi)^{N-1}-2\,\lambda \La^{6-N}\frac{1}
{\bar{\phi}\phi}\mbox{Re}\,\phi^{N}\,\ \ \ (\lambda\sim 1)\ .
\eqe
Note that adding a constant to $V$ destroys the existence of 
BPS saturated solutions and therefore property 1). Therefore, once the field $\phi$ relaxes 
to one of the $N$ vacua (see Eq.\,(\ref{vac}) the potential $V$ does not contribute to the cosmological constant! 
Modulo a constant phase $\e^{i\delta}$ 
${V}^{1/2}$ is given as
\eqb
\label{pot1/2}
{V}^{1/2}=\frac{\La^3}{\phi}-\lambda \frac{\phi^{N-1}}{\La^{N-3}}\ ,\ \ \ (\lambda\sim 1)\ . 
\eqe
The choice of phase is correlated with a choice of gauge. For the ground state solution we 
fix the gauge to shuffle as much physics as 
possible into the scalar sector. At $T>0$ we fortunately have a 
criterion on how to do this: solutions to eqs.\,(\ref{BPS}) must be {\sl periodic}. 
Only the choice $\delta=\pm\pi$ leads to periodic solutions \cite{Hof1}. 
In ref.\,\cite{HofKeil} it was shown that for sufficiently large$N$up to the first 
point of inflexion $|\phi|_c$ of the potential the positive-power part 
in ${V}^{1/2}$ can be neglected, and that in this domain the solutions are 
\eqb
\label{BPSsol}
\phi^{(n)}(\tau)=\sqrt{\frac{\La^3\beta}{2|n|\pi}}\,
\e^{2n\pi i\frac{\tau}{\beta}}\ ,\ \ (n\in{\bf Z})\ .
\eqe
It is observed that inflexion point $|\phi|_c$ does exist only for N$>$8. There is a slow convergence 
$\lim_{N\to\infty}|\phi|_c=\La$. For $N\le$8 
there is no conventional phase transition (no tachyonic fluctuations). 
However, as we shall see below, a (quasi)classical discussion of the 
transition is only appropriate in the limit of large$N$anyhow. 
The distinct topologies are labelled by $n$. Since within each given 
topology the BPS saturated solution is the 
one of lowest spatial action density these solutions are {\sl stable} 
against classical perturbations. In ref.\,\cite{HofKeil} it was shown that there exists a pure gauge configuration $A_\mu$, 
which solves the Euclidean Maxwell equations 
in a gravitional background and the {\sl background} of the Higgs field $\phi^{(1)}$ (and trivially also for $\phi^{(n)}$). 
Restricting to $n=1$, it is readily seen that the Higgs mechanism induced 
vector mass $m_A$ is much smaller than the mass $m_\phi$ of the Higgs quanta for $|\phi|<|\phi|_c$:
\eqb
\label{masses}
\frac{m_A}{m_\phi}=\frac{e}{\sqrt{6}}\,\left(\frac{|\phi|}{\La}\right)^3 \ .
\eqe
For $e<1$ there is a considerable suppression of $m_A$ as compared to 
$m_\phi$ since $|\phi|_c\le\Lambda$ \cite{HofKeil}. Therefore, the 
Born-Oppenheimer approximation, which led to the above ground state solution, is justified. 
On the other hand, using solution $\phi^{(1)}$ explicitly, we derive
\eqb
\label{mp}
\frac{m_\phi}{T}=\sqrt{6}\times 2\pi\sim 15.4\ ,\ \ \ \ \ (|\phi^1|<|\phi|_c)\ .
\eqe
So with prevailing momenta $p\sim T$ scalar 
quantum fluctuations of mass $m_\phi$ are to be neglected. They can be viewed as 
implicitly contained in $V$. But this is exactly what one 
expects from an effective potential of explicit mass scale $\La$: 
Quantum fluctuations of momenta 
larger than $\La$ are integrated out and do not 
deform the classically obtained ground state. For the vector contributions 
one may argue that at sufficiently small coupling $e$ the deformation 
of the ground state is marginal. As a consequence of the dynamics 
outlined above the ground state behaves like a medium with 
{\sl constant} specific heat $2\pi\La^3$ \cite{HofKeil}.

Let us now look at the regime where $|\phi|\sim|\phi|_c$. For large$N$
we have $|\phi|_c\sim \La$. At $|\phi|_c$ scalar fluctuation become {\sl massless}. 
Therefore they are dynamically relevant. For $|\phi|$ slightly larger 
than $|\phi|_c$ scalar fluctuations 
are {\sl tachyonic}, and hence 
they strongly drive the system towards phase overened by the (spontaneously broken) $Z_N$ symmetry. 
Gauge invariance masquerades as 
a discrete symmetry at low energy \cite{KraussWilczek}. Viewed from a Euclidean perspective, 
BPS saturated solutions still exist, but 
the modulus of $\phi$ starts to vary along the Euclidean 
time dimension \cite{Hof1}. This signals that the ground state becomes unstable and 
that one approaches a phase transition. Once $\phi$ relaxes to one of the$N$points $\phi_k$ of 
minimal energy (${V}^{1/2}=0$) with
\eqb
\label{vac}
\phi_k=\Lambda\,\exp\left[\,2\pi i\,\frac{k}{\mbox{N}}\,\right]\ ,\ \ \ \ \ \ (k=0,\cdots,\mbox{N--1})\ ,
\eqe
Euclidean and Minkowskian description are equivalent. This is also true for the description of 
topologically non-trivial defects such as $Z_N$ domain walls \cite{Hof1}. For a similar situation \cite{Linde}
it has been shown, however, that as a result 
of the rapidity of the phase transitions domains cannibalize very 
efficiently within periods of the order of $\Lambda^{-1}$. 
Due to the topological triviality of $\phi_k$ the ground state then is a perfect thermal insulator. 
Moreover, along the lines of refs.\,\cite{Cohn} it can be argued for the $Z_N$ 
symmetric theory that relic vector bosons couple very weakly to matter not 
charged under the original SU($N$) gauge group if$N$is not too small. This is a consequence of the 
absence of matter-field-composed lower dimensional operators 
which possibly could mediate a more rapid decay. 
So viewed within a cosmological setting relic vector bosons contribute to the dark matter of the universe. 
Extending the fundamental SU($N$) theory by a fundamental fermionic matter 
sector, one should assign $Z_N$ charge to color 
neutral composite operators of these fields which are 
relevant in the confining phase. The latent heat $\Delta Q$ of the transition 
between the abelian gauge theory and the center symmetric 
phase should be defined by the value of $V$ at $|\phi|_c$. It is a measure for the fourth 
power of the reheating temperature when applying the model to cosmic inflation. 
Keeping $\La$ fixed, $\Delta Q$ viewed as a function of $N$
has a maximum of $\sim 1.62\,\La^4$ at $N$=14 
and for N$\to\infty$ approaches $\La^4$. 

The mass of scalar fluctuations around one of the $N$ vacua is $m_\phi=\sqrt{2}\mbox{N}\La$. So if 
we interpret this phase as the confining one we see that even though 
quantum fluctuations should in principle matter at $|\phi|=\La$ the 
description in terms of classical fields becomes better 
and better with growing N. This may relate to the observation that bound states of 
light quarks aquire zero width in the limit N$\to\infty$ since there are no 
fluctuations in the vacuum which possibly could mediate a decay. 

\section{Summary}

To summarize, based on a small set of principles we have 
{\sl constructed} an effective potential for the description of global ground state properties 
of thermalized SU($N$) pure gauge theory in effective field variables for large N. To motivate 
the constraint on the potential to allow for BPS saturated solutions along 
a compact dimension we have constructed an example where BPS 
saturation of the dynamical monopole condensate leads to a decoupling 
of gravity for a spacetime of 4-torus topology. As one application 
this potential puts to practice the appeal of the gauge principle to cosmic evolution. In particular, 
the questions of how inflationary cosmology works, 
how it is terminated, and what the origin of ultra-high energy cosmic rays and the reason for 
their stability is may be addressed in an orderly fashion \cite{HofKeil}.

\section*{Acknowledgements}    

The author would like to thank P. van Baal, F. Bruckmann, M. Keil, and 
P. Majumdar for stimulating discussions. The warm 
hospitality and financial support during a visit to the 
Instituut-Lorentz last December are gratefully acknowleged.

\bibliographystyle{prsty}

\begin{thebibliography}{10}

\bibitem{ColemanWeinberg}
S. Coleman and E. Weinberg, Phys. Rev. {\bf D7}, 1888 (1973).\\ 
C. Ford, I. Jack, and D. R. T. Jones, Nucl. Phys. {\bf B387}, 373 (1992), Erratum-ibid. {\bf B504}, 551 (1997).\\ 
C. Ford, D. R. T. Jones, P. W. Stephenson, and M.B. Einhorn, Nucl. Phys. {\bf B395}, 395 (1993).\\ 
C. Wetterich, Phys. Lett. {\bf B301}, 90 (1993).\\ 
L. V. Laperashvili, H. B. Nielsen, and D. A. Ryzhikh, hep-th/0109023. 

\bibitem{SMT}
D. B\"odeker, W. Buchm\"uller, Z. Fodor, and  T. Helbig, Nucl. Phys. {\bf B423}, 171 (1994).\\ 
Z. Fodor and A. Hebecker, Nucl. Phys. {\bf B432}, 127 (1994).

\bibitem{KraussWilczek}
L.M. Krauss and F. Wilczek, Phys. Rev. Lett. {\bf 62}, 1221 (1989).

\bibitem{center}
G. Mack and E. Pietarinen, Nucl. Phys. {\bf B205}, 141 (1982).

\bibitem{Susskind}
E. Farhi and L. Susskind, Phys. Rept. {\bf 74}, 277 (1981).

\bibitem{LeutwylerGasser}
S. L. Glashow and S. Weinberg, Phys. Rev. Lett. {\bf 20}, 224 (1968).\\ 
S. Weinberg, Phys. Rev. {\bf 166}, 1568 (1968).\\ 
J. Gasser and H. Leutwyler,  Annals Phys. {\bf 158}, 142 (1984).\\  
J. Gasser and H. Leutwyler, Nucl. Phys. {\bf B250}, 465 (1985). 
    
\bibitem{dualsc}
G. 't Hooft, Nucl. Phys. {\bf B190}, 455 (1981).\\  
T. Suzuki, Prog. Theor. Phys. {\bf 80}, 929 (1988); {\bf 81}, 752 (1989).\\ 
S. Maedan and T. Suzuki, Prog. Theor. Phys. {\bf 81}, 229 (1989).\\ 
H. Ichie, H. Suganuma and H. Toki, Phys. Rev. {\bf D52}, 2944 (1995).\\ 
M. N. Chernodub, F. V. Gubarev, M. I. Polikarpov, 
V. I. Zakharov, Nucl. Phys. {\bf B600},163 (2001), hep-th/0010265. 

\bibitem{centerlat}
L. Del Debbio, M. Faber, J. Greensite, and S. Olejnik, Phys. Rev. {\bf D55}, 2298 (1997).\\ 
P. de Forcrand and M. D'Elia, Phys. Rev. Lett. {\bf 82}, 4582 (1999).

\bibitem{Suzuki}
S. Maedan and T. Suzuki, Prog. Theor. Phys. {\bf 81}, 229 (1989).

\bibitem{CCJ}
C. G. Callan, S. Coleman, and R. Jackiw, Annals. Phys. {\bf 59}, 42 (1970).

\bibitem{Hof4}
R. Hofmann, Phys. Lett. {\bf B520}, 257 (2001), hep-ph/0109007.\\ 
R. Hofmann, Nucl. Phys. {\bf B623}, 301 (2002), hep-ph/0109008.\\ 
K. G. Chetyrkin, S. Narison, V. I. Zakharov, Nucl. Phys. B{\bf 550}, 353 (1999).

\bibitem{BPS}
E. B. Bogomolny, Sov. J. Nucl. Phys. {\bf 24}, 449 (1976).\\ 
M. K. Prasad and C. M. Sommerfield, Phys. Rev. Lett. {\bf 35}, 760 (1975).

\bibitem{LDS}
X. Hou, A. Losev, and M. Shifman,  Phys. Rev. {\bf D61}, 085005 (2000) .\\   
G. Dvali and M. Shifman,  Phys. Lett. {\bf B454}, 277 (1999). 

\bibitem{Hof1}
R. Hofmann, Phys. Rev. {\bf D62}, 065012 (2000), hep-th/0004178.

\bibitem{HofKeil}
R. Hofmann and M. T. Keil, hep-ph/0111076.

\bibitem{Linde}
G. N. Felder, L. Kofman, A.D. Linde, Phys. Rev. {\bf D64}, 123517 (2001), hep-th/0106179. 

\bibitem{Cohn}
K. Hamaguchi, Y. Nomura, and T. Yanagida, Phys. Rev. {\bf D58}, 103503 (1998), hep-ph/9805346.\\ 
K. Hamaguchi and Y. Nomura, Phys. Rev. {\bf D59}, 063507 (1999), hep-ph/9809426.  
 

\end{thebibliography}

\end{document}